\documentclass[a4paper,12pt]{article}

\newcommand{\qinv}{q^{-1}}
\newcommand{\qana}{$q$-analogue }
\newcommand{\sutwo}{$SU(2)$ }

\newcommand{\ie}{\textit{i.e.}}

\begin{document}
%\vspace*{-2cm}
\begin{flushright}
%ver.980929-2(final)
\end{flushright}
\vspace{-9mm}

\vspace*{-.6in}
\thispagestyle{empty}

\vspace{.5in}
{\Large
\begin{center}
\textbf{A \qana of Nahm's formalism\\ for\\ self-dual gauge fields}
\end{center}}
\medskip
\begin{center}
Masaru Kamata \footnote{e-mail address: nkamata@minato.kisarazu.ac.jp}\\
\emph{Kisarazu National College of Technology\\
      Kisarazu, Chiba 292-0041, Japan}\\
\end{center}
\begin{center}
and
\end{center}
\begin{center}
Atsushi Nakamula\footnote{e-mail address: nakamula@csc.kitasato-u.ac.jp}
\\
\emph{Department of Physics, School of Science\\ Kitasato University,\\ Sagamihara, Kanagawa 228-8555, Japan}\\
\end{center}
\vspace{0.5cm}

\begin{center}
ABSTRACT
\end{center}
We present a \qana of Nahm's formalism for the BPS monopole, which gives self-dual gauge fields with a deformation parameter $q$. The theory of the basic hypergeometric series is used in our formalism. In the limit $q \to 1$ the gauge fields approach the BPS monopole and Nahm's result is reproduced.

\vfill

\begin{flushleft}
keywords: ADHM, Nahm, self dual, $q$-analogue, basic hypergeometric series
\end{flushleft}

\thispagestyle{empty}

\newpage
\setcounter{page}{2}
%\section{Introduction}
%(1)ADHMN-formalism
Instantons and monopoles are classical solutions in Yang-Mills and Yang-Mills-Higgs gauge theories, respectively.
A remarkable fact is that both of them are characterized by the self-duality.
These solitonic objects localized in $R^4$ or $R^3$, respectively,  were extensively discussed in the latter half of the 1970's both from the physical and mathematical points of view.
A method of constructing the instanton solutions was summarized as the ADHM formalism \cite{AHDM, Corri}.
The formalism was successfully applied by Nahm \cite{Nahm80-1} to the BPS monopole \cite{Bogo,PS}, which could be interpreted as a limiting case when the instanton number approaches infinity. 
Furthermore, Nahm extended his formalism to general monopole configurations \cite{Nahm82}, where famous Nahm's equations played a central role \cite{Hit83}.
There are some activities to solve Nahm's equations in  recent literature 
\cite{Mera, BrzeMera, HMM, Diaco, HoughtonSutcliffe97, HoughtonSutcliffe96-2, HoughtonSutcliffe96-1}. 
In this paper, we point out that there is another possibility in application of the ADHM and Nahm's (ADHMN) formalism to self-dual gauge fields.

%2)q-analysis in physics 
In integrable field theories and solvable lattice models, some systems preserve their solvability under certain deformations with continuous parameters.
One of them is the $q$-KZ equation, a \qana of the Knizhnik-Zamolodchikov equation \cite{KZ} in conformal field theory, derived by Frenkel and Reshetikhin \cite{FR}.
Although the differential operators in the KZ equation are replaced by the  $q$-difference operators, the system still remains to be integrable.
This suggests that we can introduce, in certain systems, a discrete space instead of continuous one without loss of some properties of original systems.
Inspired by the idea, we construct, in the following, a \qana of the ADHMN formalism for \sutwo gauge fields which preserves the self-duality, where $q$ is a continuous deformation parameter.
We show that our formalism gives a self-dual configuration which approaches the BPS monopole if we take the limit $q\rightarrow1$.
  
%\section{The construction}

Before doing our analysis, we give a brief review of the ADHMN formalism \cite{AHDM, Corri, Nahm80-1, Hit83}.
For \sutwo gauge group, the instanton associated with the 2nd Chern class of degree  $k$ is given by the vector $v$ in a $k+1$ dimensional quaternionic vector space.
For the monopole configurations, where the Higgs field can be considered the $A_0$ component of gauge fields, Nahm introduced the complex Hilbert space ${\cal L}^2[-\frac{1}{2},\frac{1}{2}]$ tensored with quaternion $H$ instead of the finite dimensional vector space.
The ADHMN formalism says that the gauge potentials $A_\mu$ which make the self-dual field strengths can be given by
\begin{equation}
A_\mu=\,<v,i\partial_\mu v>, \label{gauge p}
\end{equation}
where the symbol $<\,\,\,,\,>$ denotes an inner product over the vector spaces mentioned above. The vector $v$ is to be determined through the linear operator $\Delta$ restricted to the conditions that $\Delta^{*}\Delta$ is quaternionic real and invertible, where $\Delta^*$ is an adjoint of $\Delta$. 
Here the adjoint is in accordance with the inner product in (\ref{gauge p}).
The conditions on $v$ are:
\begin{equation}
\Delta^{*}v=0, \label{Dv}
\end{equation}
and
\begin{equation}
<v,v>\,=1, \label{norm of v}
\end{equation}
which ensures the hermiticity and the covariance under gauge transformation of $A_\mu$.
In ref.\cite{Nahm80-1}  Nahm chose the operator
\footnote{
Throughout the paper, we use the conventions: $^\dagger$ denotes hermitian conjugation, $\tau_\mu=(1,i\sigma_1,i\sigma_2,i\sigma_3)$ and $x_\mu=(x_0,x_1,x_2,x_3)$ are quaternion elements and spacetime coordinates, respectively,  and $x:=\sum_{\mu=0}^3x_\mu\tau_\mu$ and $r:=(x_1^2+x_2^2+x_3^2)^{1/2}$.
}
 $\Delta^{*}=i\partial_z+x^\dagger$ acting on the ${\cal L}^2\otimes H$ to obtain the BPS monopole configuration.

Now we shall consider the self-dual configuration associated with another class of infinite dimensional vector space, \ie, a vector space $l^2(I_q)$ tensored with $H$.
The $l^2(I_q)$ is defined on the discrete points $I_q:=\{\pm\frac{1}{2},\pm\frac{1}{2}q,\pm\frac{1}{2}q^2,\cdots\}$, $q$ being a real parameter of \, $0<q<1$. We see that there exists an accumulation point at zero.
We describe the elements of $l^2\otimes H$ as $v=f(x_\mu,z;q)$ where $z$ parametrizes $I_q$.

First of all, we have to define the linear operator $\Delta$ and $\Delta^*$ acting on $l^2\otimes H$ subject to the conditions which ensure self-dual configurations.
As we will see, the following definition to the inner product over the $l^2$ space gives the appropriate result.
We define the adjoint vector $v^*$ of $v=f(x_\mu,z;q)$:
\begin{equation}
v^*=[f(x_\mu,z;q)]^*\ =\ f^\dagger(\qinv x_\mu,qz;\qinv). \label{*}
\end{equation}
Note that we easily see $(v^{*})^{*}=v$.
Using this adjoint vector, we define our inner product:
\begin{equation}
<w,v>\,=\int^{1/2}_{-1/2}w^* v \,d_qz:= \int^{1/2}_0w^* v \,d_qz-\int^{-1/2}_0w^* v \,d_qz, \label{<,>}
\end{equation}
where the integral is so called ``Thomae-Jackson integral" defined as
\begin{equation}
\int^a_{0}f(z) \,d_qz=a(1-q)\sum^\infty_{n=0}f(aq^n)q^n, \label{qint}
\end{equation}
so that (\ref{<,>}) is in fact the infinite summation over the discrete points of $I_q$.
Hereafter we call this ``integral" $q$-integral, for simplicity. 
We now introduce the $q$-derivative operation:
\begin{equation}
D_qf(z):=\frac{f(z)-f(qz)}{(1-q)z}.\label{Dq}
\end{equation}
We can observe that our inner product makes the $q$-derivative operator $iD_q$  self-adjoint up to boundary terms, that is:
\begin{equation}
<iD_qw,v>\,=\,<w,iD_qv>, \label{selfadj}
\end{equation}
 provided that the vectors $v$ and $w$ are of the form $v(xz,x^\dagger z;q)$ and $w(xz,x^\dagger z;q)$.
If we define the linear operator $\Delta^*$ which restricts the vector $v$ in (\ref{Dv}) as 
\begin{equation}
\Delta^*=iD_q+x^\dagger,\label{Dstar}
\end{equation}
then we have $\Delta=iD_q+x$ due to the self-adjointness of $iD_q$ and the product $\Delta^*\Delta$ is quaternionic real, that is,
\begin{eqnarray}
&\Delta^*\Delta=-D_q^2+2ix_0D_q+|x|^2&\nonumber\\
     &\ \qquad=(iD_q+\rho_{+})(iD_q+\rho_{-}),&
\end{eqnarray}
where $|x|^2:=\sum_{\mu=0}^3x_\mu^2$ and $\rho_\pm:=x_0\pm ir$.
Thus if the invertible condition is fulfilled, we can obtain a self-dual gauge configuration through the vector $v$ in $l^2\otimes H$ determined by (\ref{Dv}) and (\ref{norm of v}).
We postpone the proof of the invertibility at the end of the paper.

Since the definition to inner products, in the ordinary ADHMN formalism, are based on the hermitian conjugation, we have the hermiticity and the gauge covariance of $A_\mu$ through (\ref{norm of v}).
Although our \qana of the formalism preserves the self-duality, we have $A_\mu^*=A_\mu$ instead of the hermiticity.
Further, we can also observe the covariance under ``deformed gauge transformation", which is realized by multiplying $v$ by a matrix $g(x_\mu;q)$ on the right.
Under this action, the gauge  potential transforms as:
\begin{eqnarray}
A_\mu=\,<v,i\partial_\mu v> \rightarrow <vg,i\partial_\mu vg>\,=g^*A_\mu g + g^*i\partial_\mu g,
\end{eqnarray}
thus the Yang-Mills action is invariant if $g$ is subject to $g^*=g^{-1}$.

%%%4)the derivation of eigenvector
Next, we solve (\ref{Dv}) and (\ref{norm of v}) to find the explicit  form of $v$.
For this purpose, we make use of the theory of basic hypergeometric series \cite{GR}.
The key formula is the $q$-binomial theorem:
\begin{equation}
\sum^\infty_{n=0}\frac{(a;q)_n}{(q;q)_n}z^n=\frac{(az;q)_\infty}{(z;q)_\infty},\label{qbin}
\end{equation}
where $(a;q)_n$ is the $q$-shifted factorial:
\begin{equation}
 (a;q)_n=\left\{ \begin{array}{lll} (1-a)(1-aq)(1-aq^2)\cdots(1-aq^{n-1})& 
\mbox{for} & n \ge 1, \\ 1 & \mbox{for} & n=0, \end{array}\right.
\end{equation}
and we also define
\begin{equation}
(a;q)_\infty=\prod_{m=0}^\infty(1-aq^m).
\end{equation}
We introduce a \qana of the ordinary exponential function $e^z$ defined as
\begin{equation}
e_q(z)=\sum^\infty_{n=0}\frac{z^n}{(q;q)_n}=\frac{1}{(z;q)_\infty},\label{eq}
\end{equation}
where the second equality is a consequence of (\ref{qbin}).
From the infinite series expression in (\ref{eq}), we easily obtain the formula:
\begin{equation}
D_qe_q(\alpha(1-q)z)=\alpha e_q(\alpha (1-q)z),
\end{equation}
where $\alpha$ is an arbitrary constant matrix.
Thus we find the functional form of $v$ as:
\begin{equation}
v=e_q(ix^\dagger (1-q)z)N(x_\mu;q),\label{v=eN}
\end{equation}
$N(x_\mu;q)$ being a ``normalization function" to be determined by the condition (\ref{norm of v}).
Since both $e_q$ and $N$ are quaternion-valued functions, we have to take care the order of them, in general.

We now fix the functional form of $N$.
First we observe that the $l^2$ norm of $e_q(ix^\dagger (1-q)z)$ is :
\begin{equation}
<e_q(ix^\dagger (1-q)z),e_q(ix^\dagger (1-q)z)>\,=\Lambda_+(x_0,r;q)1+\Lambda_-(x_0,r;q)\hat x,\label{e,e}
\end{equation}
where $\hat x:=\sum_{j=1}^3x_j\sigma_j/r$ and the functions $\Lambda_\pm$ are 
\begin{equation}
\Lambda_\pm(x_0,r;q)
=\frac{1-q}{2}\left\{\sum_{n=0}^\infty\frac{(\frac{\rho_+}{\rho_-};q)_{2n}}{(q;q)_{2n+1}}\biggl(i\frac{(1-q)\rho_-}{2}\biggr)^{2n}\pm (\rho_+\leftrightarrow\rho_-)\right\},\label{Apm}
\end{equation}
which are no longer quaternion valued.
Note that $\hat x$ is an hermitian matrix depending only on the angles $\theta$ and $\phi$ of the spherical coordinate system in $R^3$.
The outline of the derivation of (\ref{e,e}) and (\ref{Apm}) is as follows.
The adjoint of $e_q$ is:
\begin{equation}
[e_q(ix^\dagger(1-q)z)]^*=E_q(-ix(1-q)z),\label{estar=E}
\end{equation}
where $E_q(z)$ is another \qana of the exponential function \cite{GR} defined as
\begin{equation}
E_q(z)=\sum^\infty_{n=0}\frac{q^{n(n-1)/2}}{(q;q)_n}z^n=(-z;q)_\infty
.\label{Eq}
\end{equation}
We can prove (\ref{estar=E}) by the fact: 
\begin{equation}
(q;q)_n=(\qinv;\qinv)_n(-q)^nq^{n(n-1)/2}.
\end{equation}
The second equality in (\ref{Eq}) is a consequence of (\ref{qbin}), too.
Thus if we directly apply the $q$-binomial theorem (\ref{qbin}) to the product $E_q(-ix(1-q)z)e_q(ix^\dagger(1-q)z)$, we obtain the following formula after straightforward calculation:
\begin{equation}
E_q(-ix(1-q)z)e_q(ix^\dagger(1-q)z)=\lambda_+(x_0,r;z;q)1+\lambda_-(x_0,r;z;q)\hat x
\end{equation}
where
\begin{equation}
\lambda_\pm(x_0,r;z;q)=\frac{1}{2}\left\{\sum^\infty_{n=0}\frac{(\frac{\rho_+}{\rho_-};q)_n}{(q;q)_n}(i\rho_-(1-q)z)^n\pm\sum^\infty_{n=0}\frac{(\frac{\rho_-}{\rho_+};q)_n}{(q;q)_n}(i\rho_+(1-q)z)^n\right\}.
\end{equation}
Here we used the following fact:
\begin{eqnarray}
x^m=\frac{1}{2}(\rho_+^m+\rho_-^m)1+\frac{1}{2}(\rho_+^m-\rho_-^m)\hat x,\\
x^{\dagger m}=\frac{1}{2}(\rho_+^m+\rho_-^m)1-\frac{1}{2}(\rho_+^m-\rho_-^m)
\hat x,
\end{eqnarray}
for positive integer $m$.
Performing the $q$-integration,
\begin{equation}
\int^{1/2}_{-1/2}z^nd_qz=\left\{ \begin{array}{lll} 0& \mbox{for odd}& n, \\ 
\frac{1-q}{1-q^{n+1}}\frac{1}{2^n}& \mbox{for even}& n, \end{array}\right.
\end{equation}
for positive integer $n$, we thus obtain the result (\ref{e,e}) and (\ref{Apm}). Next, we consider the normalization condition (\ref{norm of v}), which now turns out to be:
\begin{equation}
N(x_\mu;q)^*(\Lambda_+1+\Lambda_-\hat x)N(x_\mu;q)=1.
\end{equation}
We can easily see that the inverse square root matrix $(\Lambda_+1+\Lambda_-\hat{x})^{-1/2}$ works well for $N$ due to the fact $\Lambda_\pm^*=\Lambda_\pm$, and which gives the result:
\begin{eqnarray}
N(x_\mu;q)&=\frac{1}{2}\{(\Lambda_++\Lambda_-)^{-\frac{1}{2}}+(\Lambda_+-\Lambda_-)^{-\frac{1}{2}}\}1&\nonumber\\
&+\frac{1}{2}\{(\Lambda_++\Lambda_-)^{-\frac{1}{2}}-(\Lambda_+-\Lambda_-)^{-\frac{1}{2}}\}\hat x&, \label{N}
\end{eqnarray}
and we have $N^*=N$, trivially.
Our derivation therefore fixes the functional form of $v$ up to the ``gauge transformation" mentioned earlier.

%5)the classical limit
We consider the limit $q \to 1$, which has to agree with Nahm's result for the BPS monopole, because the limit gives the $q$-derivative operator (\ref{Dq}) the  ordinary differential one.
 In the limit we have 
\begin{equation}
\lambda_\pm(x_0,r;z;q)\to \frac{1}{2}\left\{\sum^\infty_{n=0}\frac{(1-\frac{\rho_+}{\rho_-})^n}{n!}(i\rho_-z)^n\pm\sum^\infty_{n=0}\frac{(1-\frac{\rho_-}{\rho_+})^n}{n!}(i\rho_+z)^n\right\}
\end{equation}
then we find
\begin{equation}
 E_q(-ix(1-q)z)e_q(ix^\dagger(1-q)z) \to \cosh(2rz)+\hat{x}\sinh(2rz)
\end{equation}
which is $x_0$-independnt and leads to 
\begin{equation}
N^{-2}=\frac{\sinh{r}}{r}.
\end{equation}
Thus we can reproduce the BPS monopole as a limiting case of the deformation parameter $q$.

%5)invertibility
Finally, we show that $\Delta^*\Delta$ is invertible, which guarantees the self-duality of the gauge field configuration constructed from $v$ of (\ref{v=eN}) and (\ref{N}).
It is sufficient to find a function $F(x_\mu;z,z';q)$ such that
\begin{equation}
\Delta^*\Delta F(x_\mu;z,z';q)=\delta_q(z,z'),\label{DDinv}
\end{equation}
where the right hand side is a function which takes a non-zero value only at $z=z'$.
Since we are dealing with discrete space, the function $\delta_q(z,z')$ should be proportional to  Kronecker's delta $\delta_{z,z'}$.
Further, when we consider the limit $q\to1$, it should become the ordinary $\delta$-function in accordance with Nahm's result.
We introduce a sign or step function defined by:
\begin{equation}
\epsilon(z-z')=\left\{
	\begin{array}{ll}
	+1 & \mbox{if} \ z \ge z' \\
	-1 & \mbox{if} \ z < z' 
	\end{array}
\right. \; \mbox{for}\ 0<z',\; \mbox{and}\,
\left\{
	\begin{array}{ll}
	+1 & \mbox{if} \ z > z' \\
	-1 & \mbox{if} \ z \leq z' 
	\end{array}
\right. \; \mbox{for}\; z'<0.
\end{equation}
We can easily find that the $q$-derivative of the $\epsilon(z-z')$ with respect to $z$ gives Kronecker's delta, \ie,
\begin{equation}
D_q\epsilon(z-z')=\frac{2}{(1-q)|z'|}\delta_{z,z'}.\label{kronecker}
\end{equation}
We make use of the right hand side as the function $2\delta_q(z,z')$ in (\ref{DDinv}).
Then we obtain the solution to (\ref{DDinv}):
\begin{eqnarray}
&F(x_\mu;z,z';q)=\frac{1}{4r}\epsilon(z-z')\{E_q(-i\rho_+(1-q)qz')e_q(i\rho_+(1-q)z)\qquad&\nonumber\\
&\qquad -E_q(-i\rho_-(1-q)qz')e_q(i\rho_-(1-q)z)\}+F_0(x_\mu;z,z';q), &\label{FF}
\end{eqnarray}
where $F_0(x_\mu;z,z';q)$ is a kernel of $\Delta^*\Delta$ to be determined by boundary conditions, which we do not need to fix here. We thus have proven the invertible condition. Note that if we take the limit $q\to1$, the function $F$ becomes,
\begin{eqnarray}
F&\rightarrow&\frac{1}{4r}\epsilon(z-z')\left(e^{i\rho_+(z-z')}-e^{i\rho_-(z-z')}\right)\nonumber\\
&=&-\frac{1}{2r}e^{ix_0(z-z')}\sinh r|z-z'|,
\end{eqnarray}
where we have omitted the term including $F_0$; this is exactly Nahm's result.

%\section{Conclusions}

%6)convergent radius
In conclusion, we have constructed a \qana of the ADHMN formalism and obtained the vector $v$ from which a self-dual field strength can be made.
At present, our understanding about the physical aspects of our gauge field configuration is insufficient.
Some remaining questions are in order.
First, the gauge field configuration  will have a finite radius of convergence,  because the inner product (\ref{e,e}) is normalizable for $|x|<2/(1-q)$.
We can see this through another expression of $\Lambda_\pm$,  that is,
\begin{eqnarray}
\Lambda_\pm(x_0,r;q)=\frac{1}{2}\, {_2}\phi_1\left[
      \begin{array}{c}
        \begin{array}{cc}
          (\frac{\rho_+}{|x|})^2, & (\frac{\rho_+}{|x|})^2q
        \end{array}\\
       q^3
      \end{array}
;q^2,-\frac{(1-q)^2}{4}\rho_-^2
      \right] 
      \pm (\rho_+ \leftrightarrow \rho_-) \nonumber\\
\end{eqnarray}
where 
\begin{equation}
{_2}\phi_1\left[
      \begin{array}{c}
        \begin{array}{cc}
          a, & b
        \end{array}\\
       c
      \end{array}
;q, u
      \right]=\sum_{n=0}^\infty\frac{(a;q)_n(b;q)_n}{(q;q)_n(c;q)_n}u^n,\label{phi21}
\end{equation}
is Heine's \qana of Gauss' hypergeometric series \cite{GR}, which is known to be absolutely convergent for $|u|<1$ provided that $0<q<1$.
However, since the analytic continuation formula of (\ref{phi21}) to $|u|>1$ is known, we have a possibility to define the gauge field configuration outside the convergence region, similarly.
Second, the definition to the characteristic class of our  configuration is not clear if the convergence radius is finite.
Analogous to general monopole configurations, of which the associated vector space is ${\cal L}^2$, our configuration seems to have infinite instanton number because of the fact that the vector space $l^2$ is infinite dimensional.
Thirdly, we have observed that the gauge potential transforms covariantly under the action of a matrix $g$ subject to $g^*=g^{-1}$, which becomes \sutwo element when $q\to1$.
What is this ``deformed" transformation in our formalism?

We shall discuss these aspects of the gauge fields in a forthcoming paper.

\end{document}